\documentclass[preprint,aps]{revtex4}
\usepackage{graphicx}
\usepackage{dcolumn}

\begin{document}

\title{Stages of Relaxation of Patterns and the Role of Stochasticity on the Final Stage}

\author {Shaowen Hu}
\affiliation {Department of Chemistry,
                University of Houston,
                Houston, TX 77204}
\author {Daniel I. Goldman}
\affiliation {Center for Nonlinear Studies and Department of Physics,
              University of Texas at Austin,
              Austin, TX 78712}
\altaffiliation{Present address: Department of Integrative Biology,
                The University of California at Berkeley,
                Berkeley, CA 94720-3140}             
\author {Donald J. Kouri}
\affiliation {Departments of Chemistry, Physics and Mathematics,
                University of Houston,
                Houston, TX 77204}
\author {David K. Hoffman}
\affiliation {Department of Chemistry and Ames Laboratory,
                Iowa State University,
                Ames, Iowa 50011}
\author {Harry L. Swinney}
\affiliation {Center for Nonlinear Studies and Department of Physics,
              University of Texas at Austin,
              Austin, TX 78712}
\author {Gemunu H. Gunaratne}
\affiliation {Department of Physics,
                University of Houston,
                Houston, TX 77204}
\altaffiliation[Also at ] {The Institute of Fundamental Studies,
                Kandy 20000, Sri Lanka}
\email {gemunu@uh.edu}

\begin{abstract}

The disorder function formalism [Gunaratne et.al., Phys. Rev. E, {\bf 57}, 5146 (1998)]
is used to show that pattern relaxation in an experiment on a vibrated layer of brass beads
occurs in three distinct stages. During stage I, all lengthscales associated with 
moments of the disorder grow at a single universal rate, given by $L(t) \sim t^{0.5}$.
In stage II, pattern evolution is non-universal and includes a range of growth 
indices. Relaxation in the final stage is characterized by a single, non-universal index.
We use analysis of patterns from the Swift-Hohenberg equation to argue that mechanisms
that underlie the observed pattern evolution are linear spatio-temporal dynamics (stage I),
non-linear saturation (stage II), and stochasticity (stage III).

\end{abstract}
\pacs{PACS number(s) }

\maketitle

\section {Introduction}

Analysis of structures far from equilibrium and their evolution has necessitated the development and 
implementation of several new techniques~\cite{croAhoh}. The underlying physical systems
are dissipative and typically exhibit sensitive dependence on initial states~\cite{egoAgre};
i.e., differences between nearly identical patterns increase in time.  Further,
although structures formed from distinct initial states differ in detail, they appear to
share features that depend only on control parameters of the physical system.
These observations led Prigogine and collaborators to propose the use of irreversible 
statistical analysis using ensembles of solutions of a 
system~\cite{petApri1,petApri2,hasAsap, antAtas}. However, it is a non-trivial task
to identify measures appropriate to analyze a given class of structures.

We consider properties of large aspect 
ratio patterns that are formed via spontaneous symmetry 
breaking from a noisy, but otherwise structureless initial state. Patterns are 
generated when an external control parameter crosses a threshold, initiating a
local bifurcation to a periodic array; these arrays are typically, striped, square, triangular or 
hexagonal~\cite{croAhoh}. However, due to translational and rotational invariances of the 
underlying system, the location and orientation of these local patches are not 
synchronized over the entire domain. As they grow, these patches compete with each 
other to produce complex patterns seen in experimental and model 
systems~\cite{ouyAswi,heuAgol,bodAdeb,rose,melAumb,best,eldAvin}, see 
Figures~\ref{snapshot0}, \ref{snapshot_exp} and \ref{snapshot}. These
structures, whose details are sensitively dependent on the initial state, can be 
considered to belong to the ensemble of patterns.

Assume that such a spatio-temporal pattern can be represented by a scalar field $u({\bf x},t)$.
Several functionals of $u({\bf x}, t)$ have been proposed as measures to be used in
statistical descriptions. Most of these are
modeled on related quantities from equilibrium statistical mechanics.
The width of the structure factor $S(t)$~\cite{schAall,eldAvin,croAmei} 
(i.e., the width of the peak in the azimuthal average of 
$\langle\hat u({\bf k},t) \hat u(-{\bf k},t)\rangle)$ is one of the more frequently
studied examples.  It has been shown that the evolution of a pattern from a noisy
initial state consists of two distinct stages~\cite{schAall}: $S(t) \sim t^{-\frac{1}{2}}$ 
is obeyed until the peak amplitude of the field $u({\bf x}, t)$ saturates, beyond
which time the pattern coarsens and the decay becomes slower. For
the Swift-Hohenberg equation, Elder et al.~\cite{eldAvin}, Cross and
Meiron~\cite{croAmei}, and Hou et al.~\cite{houAsas} found that during
this second stage $S(t)$ decreased as $t^{-0.20}$ when the dynamics 
was noise free and at a faster rate $t^{-0.25}$ otherwise. 
Additional quantities have been introduced to describe 
the second relaxation region. Hou et al.~\cite{houAsas} measured cumulative domain wall
length as a function of time. For noise-free (resp. noisy) dynamics, it was shown to decay
as $t^{-0.25}$ (resp. $t^{-0.30}$). Cross et al.~\cite{croAmei} showed that a stripe orientation
correlation field decayed as $t^{-0.24}$. 

Notice that in the second relaxation stage, distinct measures (which probe multiple features
of a pattern) decay at different rates. A systematic way to study such variations is to 
define a field to quantify the ``local"  disorder, and to use multiple moments 
of this field to construct a family of characteristics to describe the dynamics 
of various lengthscales associated with the pattern~\cite{ruel}. 
Such a class appropriate for labyrinthine patterns, referred to as 
the disorder function, has been introduced~\cite{gunAjon,gunArat}. It is defined as 
\begin{equation}
\delta(\beta, t) = \left[ \frac{(2-\beta)}{({\int d^2x})}
                    \frac{\int d^2x |(\triangle + k^2) u({\bf x}, t)|^{\beta}}
                         {\langle |u({\bf x}, t)|\rangle^{\beta}} \right] {\bigg/} k^{2\beta},
\label{defn}
\end{equation}
where $k_0$ is the wavevector associated with the local planform,
$\langle|u({\bf x}, t)|\rangle$ denotes the mean of $|u({\bf x}, t)|$,
${\int d^2x}$ is the area of the pattern, and 
$\delta(\beta, t)$ ($0\le \beta < 2$) has been normalized to be scale
invariant. In this definition, the field used to quantify the local disorder 
is the magnitude $ |(\triangle + k^2) u({\bf x}, t)|$
of the Helmholtz operator. The integrand is non-zero at locations where the
local wave-number deviates from its global mean $k$, and those where the planforms 
are curved. (The latter can be seen by
considering a region of stripes with a radius of curvature $r$; i.e.,
$u({\bf x}, t) \sim exp (i k r)$. There 
$|(\triangle + k^2) u({\bf x}, t)| \sim k / r$.) 
Distinct ``moments'' $\beta$ can be used to emphasize particular features of 
a pattern; for example, large values of $\beta$ provide a higher weight to 
defects~\cite{gunAjon,golAsha}. Methods to calculate $\delta(\beta, t)$ 
for (noisy) experimental patterns have been introduced~\cite{hofAgun}.
In this paper, we will describe spatio-temporal dynamics using
$\delta^{\prime} (\beta, t) = \delta(\beta, t)^{1/\beta}$.
As can be easily seen, for each $\beta$, $\delta^{\prime} (\beta, t)$ 
is the ratio of a wave number associated with the moment $\beta$ of the 
Helmholtz operator and $k$.

The disorder function analysis of the evolution of patterns under model 
spatio-temporal dynamics from an initially noisy state reveals the  presence of 
two stages~\cite{gunArat}, consistent with the analysis of $S(t)$~\cite{eldAvin}. 
It is found that during the first stage spatio-temporal dynamics is essentially linear, and
$\delta^{\prime} (\beta, t) \sim t^{-\sigma_I}$, where $\sigma_I = \frac{1}{2}$ 
is a universal exponent~\cite{gunArat}. Thus, the wave number associated with each moment
$\beta$ reduces at a rate $t^{-\sigma_I}$, and the corresponding lengthscale
grows as $t^{\sigma_I}$. The second stage, domain-coarsening, exhibits more complex
behavior. The relaxation is described by $\delta^{\prime} (\beta, t) \sim
t^{-\sigma_{II}(\beta, t)}$~\cite{golAsha}. Unlike $\sigma_I$,
$\sigma_{II}(\beta, t)$ is model dependent; i.e., the corresponding relaxation
is not universal, and wave numbers associated with multiple moments decay with 
rates that depend on $\beta$. 

These conclusions were recently
validated in a set of experiments on pattern formation in a vibrated layer of brass
beads~\cite{golAsha}. However, the experiments showed, in addition, that for 
very long times the patterns reach a third stage; here, the indices 
$\sigma_{III} (\beta, t)$ are independent of $\beta$, but their common value is not universal.
The work reported in this paper shows that the presence of this final stage 
in systems far from the onset of patterns is one observable consequence of stochasticity 
in the underlying spatio-temporal dynamics. The results complement recent 
demonstrations on effects of noise in systems close to the onset of 
patterns~\cite{ohAahl,golAswi}.

\section {Model System and Analysis}

The model system used in our studies is an extension of the Swift-Hohenberg
equation~\cite{croAhoh}, which describes the dynamics of a scalar field
$u({\bf x},t)$ via
\begin{equation}
\label{she}
 \frac{\partial u}{\partial t} = D \Bigl(\epsilon-(\triangle+k_0^2)^2 \Bigr)u
  -u^3 - \nu (\nabla u)^2 + \eta({\bf x},t).
\end{equation}
$D$ sets the rate of diffusion, $\epsilon$ is the distance from pattern onset, 
$\nu$ is the strength of a non-variational term~\cite{gunArat}, and
$k_0$ is the wave number of the underlying planform. 
$\eta$ is a stochastic term such that 
$\langle \eta({\bf x},t) \eta({\bf x'},t') \rangle =  F \delta({\bf x}-{\bf x'})\delta(t-t')$, 
$F$ being the strength of the noise; $\eta\ne 0$ and $\eta = 0$ represent
stochastic and noise-free spatio-temporal dynamics. For suitable 
control parameters, random initial states
(whose amplitude is chosen to be much smaller than the saturated peak amplitude
of the field) evolve to patterns under Eqn.~(\ref{she}).
Several snapshots from such an evolution under noise-free
dynamics are shown in Figure~\ref{snapshot0} \cite{footnote0}.

The behavior of $\delta(1.0,t)$ corresponding to this spatio-temporal dynamics 
is shown in Figure~\ref{delta}. Statistical variation between values of $\delta(1.0, t)$
in nominally identical patterns--i.e., those with same control parameters, but different initial 
states-- is negligible~\cite{gunArat}. $\delta(1.0,t)$ exhibits an initial rapid decay followed by 
a much slower relaxation. During stage I, $\langle |u(x,t)|\rangle$
is significantly smaller than its value at large times; hence, spatio-temporal
dynamics can be approximated by the linear partial differential equation
\begin{equation}
\label{shlin}
 \frac{\partial u}{\partial t} = \bigl(\epsilon-(\triangle+k_0^2)^2 \bigr)u.
\end{equation}
The Fourier transform of $u({\bf x},t)$ satisfies 
\begin{equation}
  \hat u ({\bf k}, t) = \hat u ({\bf k}, 0) \ exp (\epsilon t - (k_0^2 -{\bf k}^2)^2 t).
\label{solution}
\end{equation}
For a given noisy initial state, this solution can be used to evaluate integrals
in Eqn.~(\ref{defn}). It is found that $\langle |u(x,t)|\rangle \sim e^{\epsilon t} t^{-1/4}$ 
and (for values of $t$ that are not very small) 
$\int d^2x |(\triangle + k_0^2) u({\bf x}, t)|^{\beta} 
\sim e^{\beta \epsilon t} t^{-3\beta /4}$~\cite{footnote1}.  Consequently, 
$\delta (\beta, t) \sim t^{-\beta/2}$ and $\sigma(\beta, t) = \sigma_I = 1/2$.

\begin{figure}
\includegraphics[width=4in]{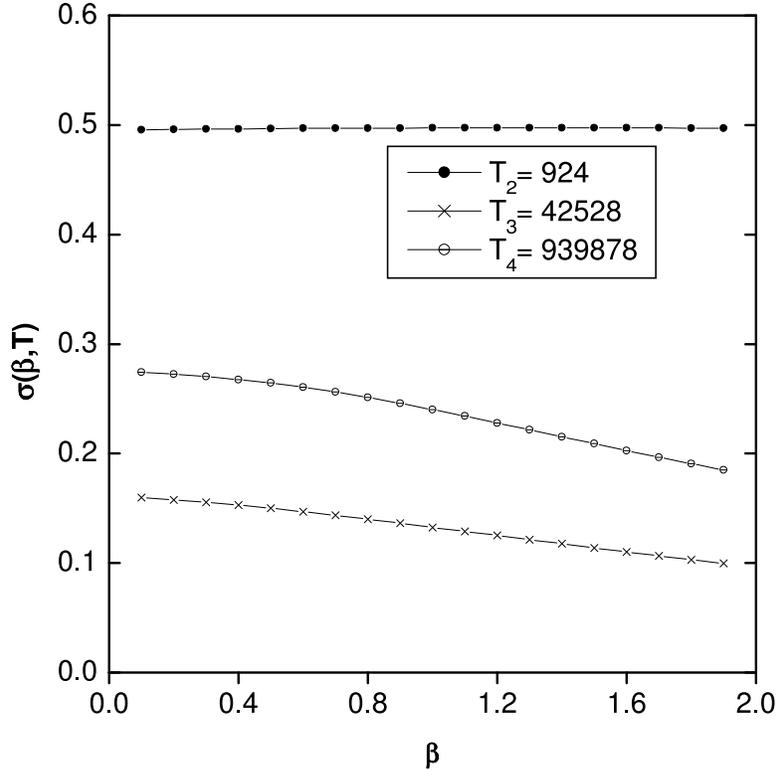}
\caption{The behavior of the function $\sigma (\beta, T)$. During the domain forming
stage $\sigma(\beta, T) = 1/2$, while it is $\beta$ dependent in the domain coarsening stage.
As in the example shown, $\sigma (\beta, T)$ also varies with time during the second stage.}
\label{sigma0}
\end{figure}

Results shown in Figure~\ref{delta} can be used to compute $\sigma(\beta, t)$. At a given
time $t$, we use the five nearest points ($t$ and the closest two points on both sides) to calculate the slope of the curves 
$\log \delta(\beta, t)$ vs. $\log t$ using a least squares fit. As shown in
Figure~\ref{sigma0}, this analysis confirms that $\sigma_I (\beta, t) = 1/2$ during stage I.

The onset of stage II coincides with the saturation of the peak value of 
the field $u({\bf x}, t)$ and beyond this time the relaxation of $\delta (1.0, t)$ is slower.
Further, as seen from  Figure~\ref{sigma0}, $\sigma(\beta,t)$ varies both with $t$
and $\beta$. The variation with $\beta$ implies that the wave numbers 
associated with $\delta^{\prime} (\beta, t)$ 
decay at rates which depend on $\beta$; i.e., lengthscales associated with multiple 
features of these patterns grow at different rates~\cite{golAsha}. 
The spatio-temporal dynamics shown in 
Figure~\ref{snapshot0} retains this feature even for very large times, when the pattern consists
of a few large domains.

\begin{figure}
\includegraphics[width=4in]{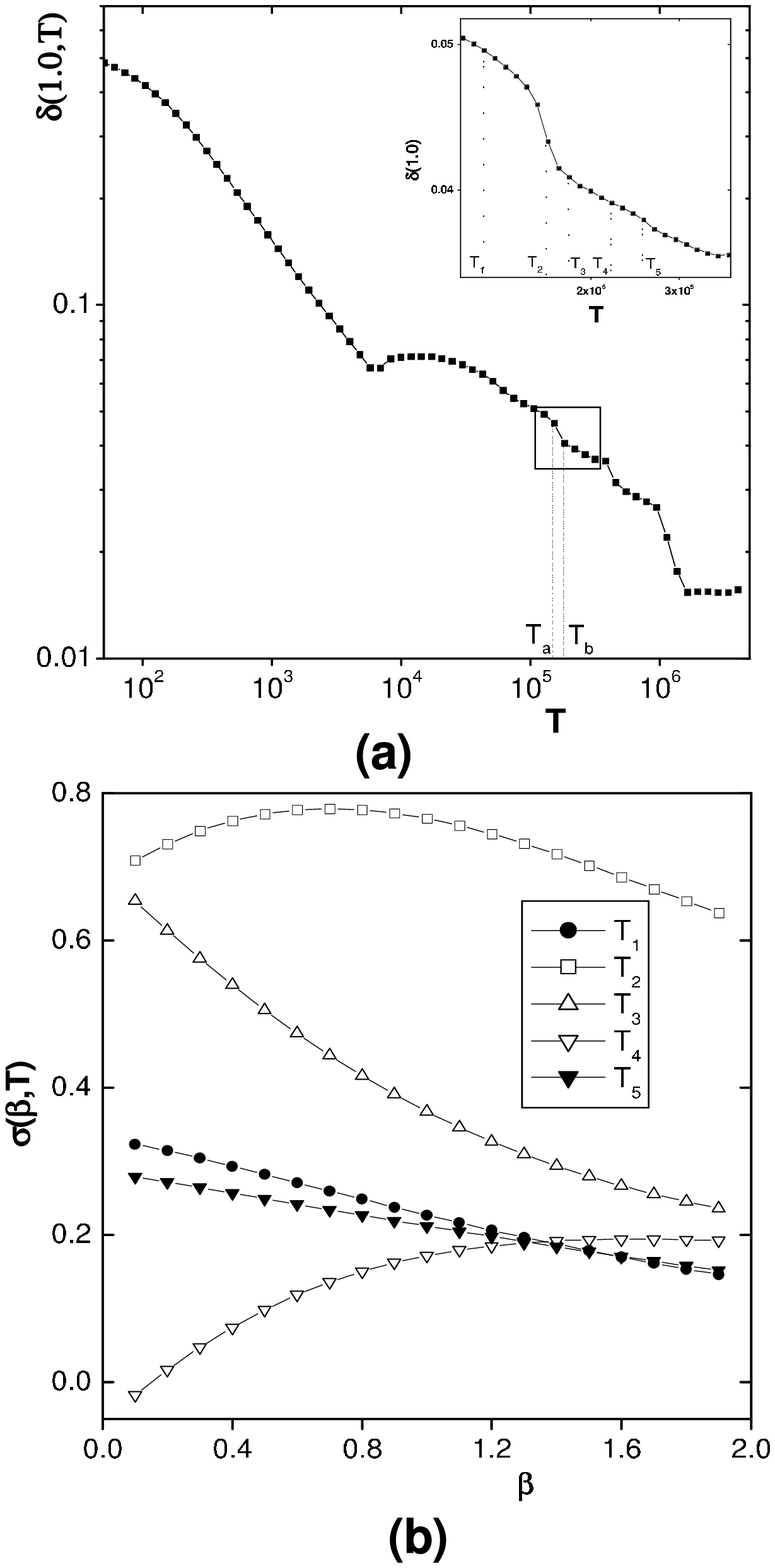}
\caption{The behavior of $\delta(\beta, t)$ and $\sigma(\beta, t)$ during the 
removal of a small domain. Notice that, while $\sigma(\beta, t)$ decreases 
monotonically before the event and following the relaxations of the neighborhood,
it exhibits more complex behavior during the event.}
\label{delta_def}
\end{figure}

Typically, $\sigma (\beta, t)$ is a monotonically decreasing function of $\beta$ during stage II;
thus, the mean distance between defects ($\beta \rightarrow 2$~\cite{gunAjon}) grows at a slower 
rate than (for example) the mean curvature or domain size.
Exceptions occur during the removal of defect(s) or small domains
from a pattern. These events can be identified by a sudden rapid decrease in
$\delta (\beta, t)$, an example of which is seen near $T_4$ in Figure~\ref{delta}.
Figure~\ref{defects} shows the pattern prior to and following the absorption of 
a small domain into a larger one. The behaviors of $\delta (1.0,t)$ and $\sigma (\beta, t)$
during this metamorphosis is shown in Figure~\ref{delta_def}(a) and (b). It can be
seen that $\sigma (\beta, t)$ which is monotonic before and after the change 
takes on a more complex form during the transition. This method can be used to identify
domain disappearance even when it is difficult to recognize the change visually (as often 
happens in patterns with a large number of domains).

\section{Results from the Experiment}

The pattern forming experiments were conducted with 0.165 mm bronze
spheres contained in a vertically oscillated circular container with a
diameter of 140 mm~\cite{melAumb,golAsha}. The layer is four particle diameters
deep, and the cell is evacuated to 4 Pa so that hydrodynamic
interaction between the grains and surrounding gas is negligible.  The
control parameters are the frequency $f$ of the sinusoidal
oscillations and the peak acceleration of the container relative to
gravity, $\Gamma=(2 \pi f)^2 A^2/g$, where $A$ is the amplitude of the
oscillation and $g$ is the gravitational acceleration. As $f$ and
$\Gamma$ are changed, a variety of temporally subharmonic patterns
including locally square, striped, or hexagonal patterns are
observed~\cite{melAumb}. The textures analyzed consisted of patterns with
square planforms~\cite{golAsha}.

The granular surface is visualized by illuminating the cell using 
a ring of LEDs surrounding the cell and is strobed at the drive frequency of
the container. The light is incident at low angles and the scattering
intensity is a nonlinear function of the height of the layer;
scattering from peaks (valleys) creates bright (dark) regions.This
intensity field is used to represent the structures~\cite{golAsha}.

Figure~\ref{snapshot_exp} shows four snapshots during the evolution of an
experimental pattern, and Figure~\ref{delta_exp} shows the corresponding 
behavior of $\delta(1.0, t)$. Stages I and II, analogous to those seen in 
patterns of the model system, can be observed. It was confirmed that the
onset of stage II corresponded to the saturation of the peak amplitude~\cite{golAsha}.
Using the leftmost interval shown in Figure~\ref{delta_exp}, the 
evolution during stage I was shown to be consistent with $\sigma (\beta, t) = 1/2$,
although there were insufficient data points for the result to be
conclusive~\cite{golAsha}. Beyond this stage, the relaxation was 
much slower, and $\sigma (\beta, t)$ depends on $\beta$~\cite{golAsha},
see Figure~\ref{expt}.  However, it was observed that at much longer times 
(e.g., during the rightmost interval in Figure~\ref{delta_exp})
the patterns reach a stage where $\sigma (\beta, t)$ is independent of $\beta$,
and its value is not the universal 1/2 of the first stage, 
see Figure~\ref{expt}. 

\section{Spatio-temporal Dynamics under Stochastic Equations}

Behavior analogous to stage III in experimental patterns is not observed in
the spatio-temporal dynamics such as that shown in Figure~\ref{snapshot0}, even at very 
large times when the pattern consists of a few domains.
Clearly, one or more ingredients that are necessary for the description 
is missing from the model we have used. Candidates include non-variational terms,
finite size effects and stochastic terms. In particular, only finite subdomains 
of the experimental system are analyzed (in contrast the model system has periodic boundary conditions), and
stochasticity has been shown to play a role in the dynamics near the onset of 
patterns~\cite{golAswi}. 

Integration of the model system shows that neither the addition of non-variational 
terms ($\nu \ne 0$) to the spatio-temporal dynamics nor conducting the analysis on
smaller sub-domains give a transition away from stage II, even at very large times.
On the other hand,  we find that the addition of stochastic terms does lead to 
such a transition, and that the new stage has characteristics described in the
last section. 

\begin{figure}
\includegraphics[width=4in]{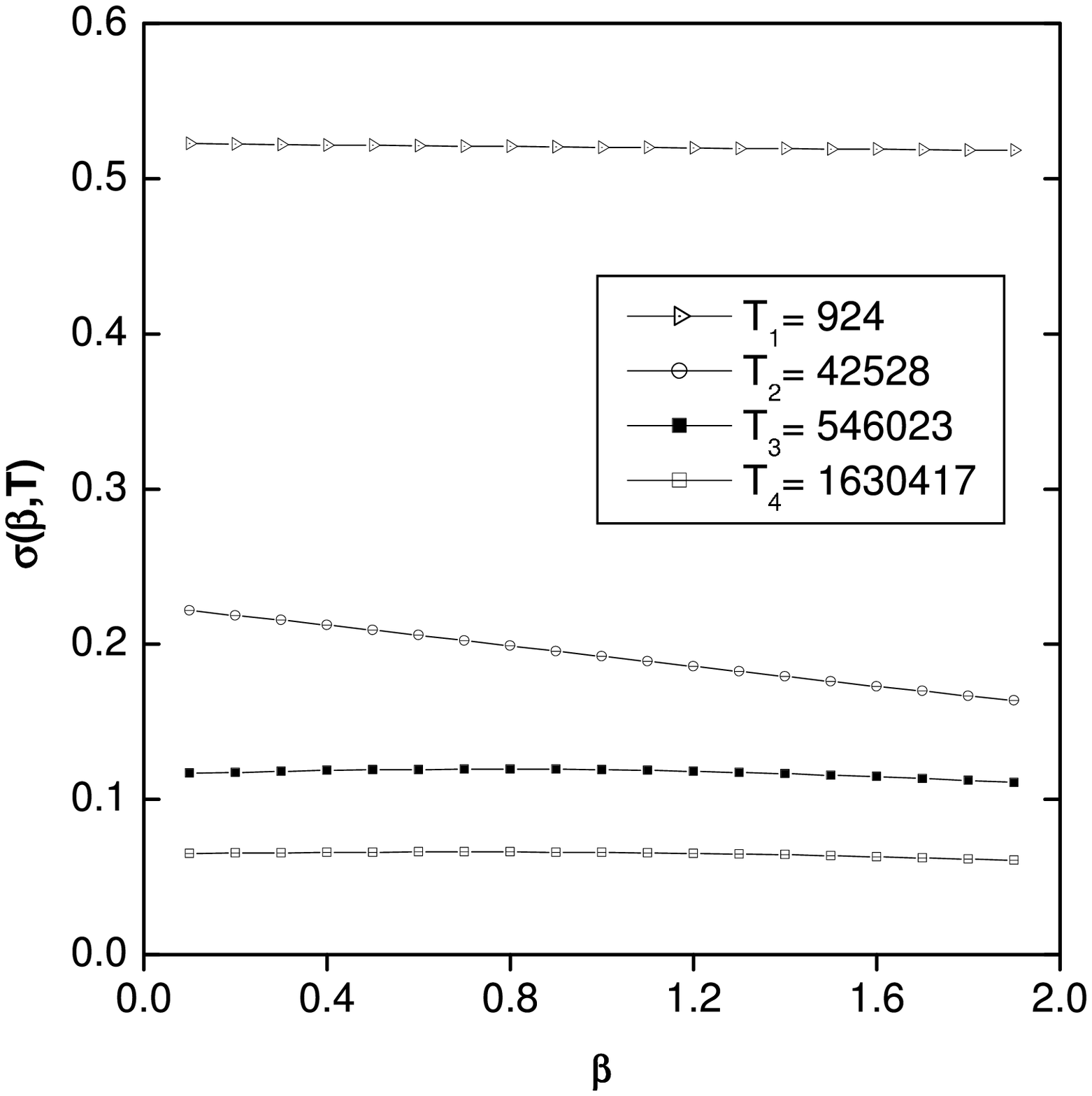}
\caption{The behavior of $\sigma (\beta, T)$ for stochastic spatio-temporal dynamics.
Observe the presence of three distinct stages, as in experimental patterns.}
\label{sigma}
\end{figure}

Figure~\ref{snapshot} shows four snapshots during the development of such a 
structure under the Swift-Hohenberg equation with stochastic terms (i.e., $F\ne 0$). 
The behavior of $\delta(1.0,t)$ is qualitatively similar to
Figure~\ref{delta}. The principal difference that can be gleaned from the
analysis of a single moment $\delta(\beta, t)$ is an increase in the decay rate,
particularly in stage II. $\sigma(\beta, t)$  
exhibits features that are similar to the corresponding noise-free dynamics;
it is independent of $\beta$ in stage I, but is a function of
$\beta$ following saturation of the field $u({\bf x}, t)$. However, at
very large times, $\sigma(\beta, t)$ is once again seen to be independent of $\beta$,
taking a value that depends on control parameters, see Figure~\ref{sigma}.
It is also found that the onset of stage III
advances with increasing intensity of the stochastic terms. For sufficiently
large values of $F$ the dynamics can move directly from stage I to III.

The implication of the $\beta$ independence of $\sigma _{III}$ is that all 
lengthscales associated with the pattern grow at the same rate during stage III.
This assertion can be tested by studying the relaxation of the width of the 
structure factor and the defect density. For the analysis, we used averages for 
four nominally identical runs with parameters given in Figure~\ref{snapshot}.
The mean decay rate of the width of the structure factor at time $T_3$ of 
Figure~\ref{sigma} is found to be $0.119 \pm 0.014$, in agreement with 
$\sigma_{III} (T_3)$. Secondly, the mean number
of defects at $T_3$ (XX HOW WAS THIS CALCULATED?) is found to decrease like $T^{\chi}$, where 
$\chi=-0.201 \pm 0.019$. The associated lengthscale (i.e., the mean distance
between defects) grows with an index  $\frac{1}{2} \chi$, once again consistent with
the value of $\sigma_{III} (T_3)$. 

\section {Discussion and Conclusions}

\begin{figure}
\includegraphics[width=5.5in]{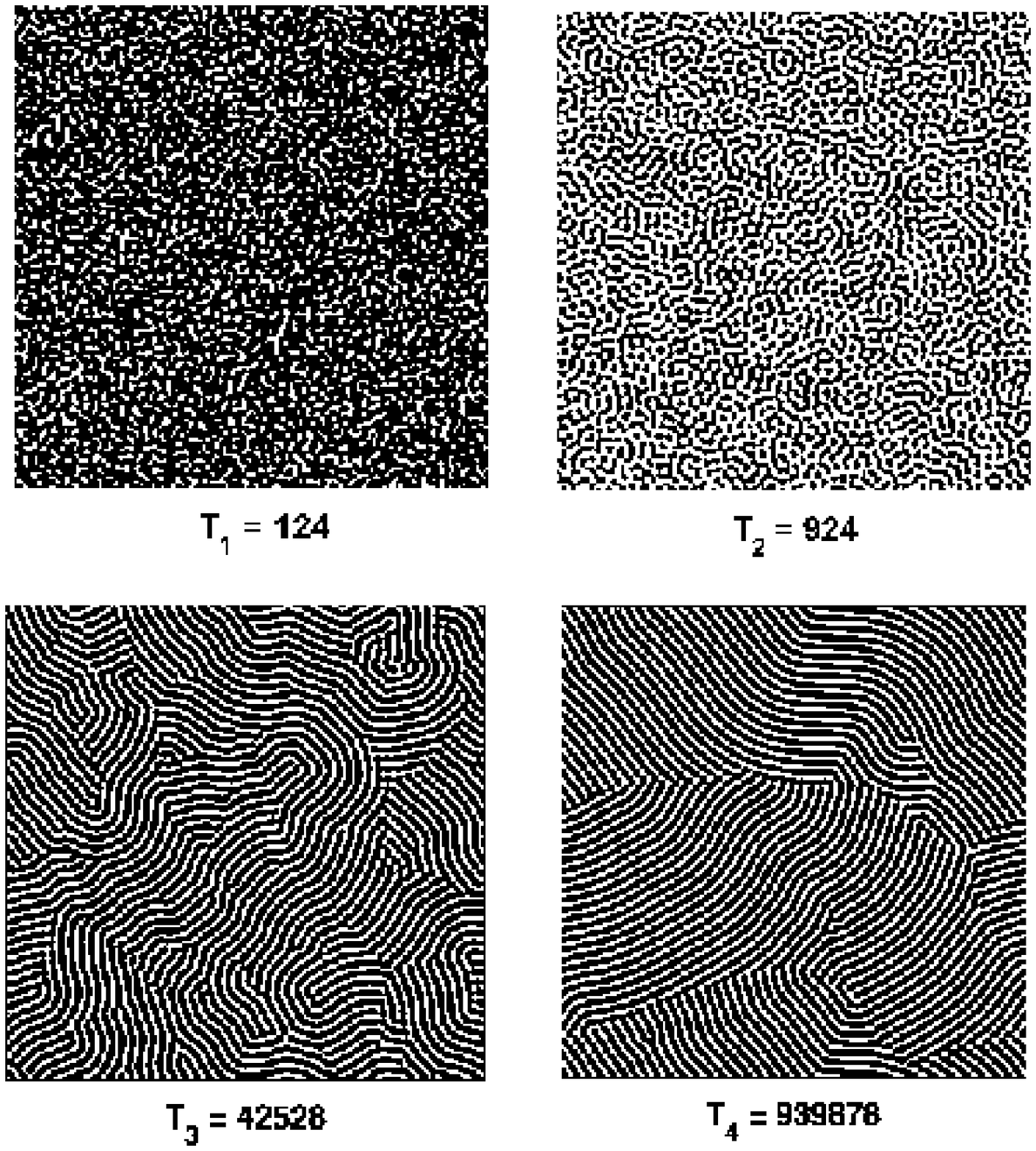}
\caption{Four snapshots of the spatio-temporal dynamics of a random initial state
($|u({\bf x}, t)| < 0.001$) under Eqn.~\ref{she} with $D=0.01$, $\epsilon=0.1$, $k_0=3$,
and $F=0$. Times $T_1=124$ and $T_2=924$ are in stage I, and $T_3=42,528$ and 
$T_4=939,878$ are in stage II.}
\label{snapshot0}
\end{figure}

\begin{figure}
\includegraphics[width=4in]{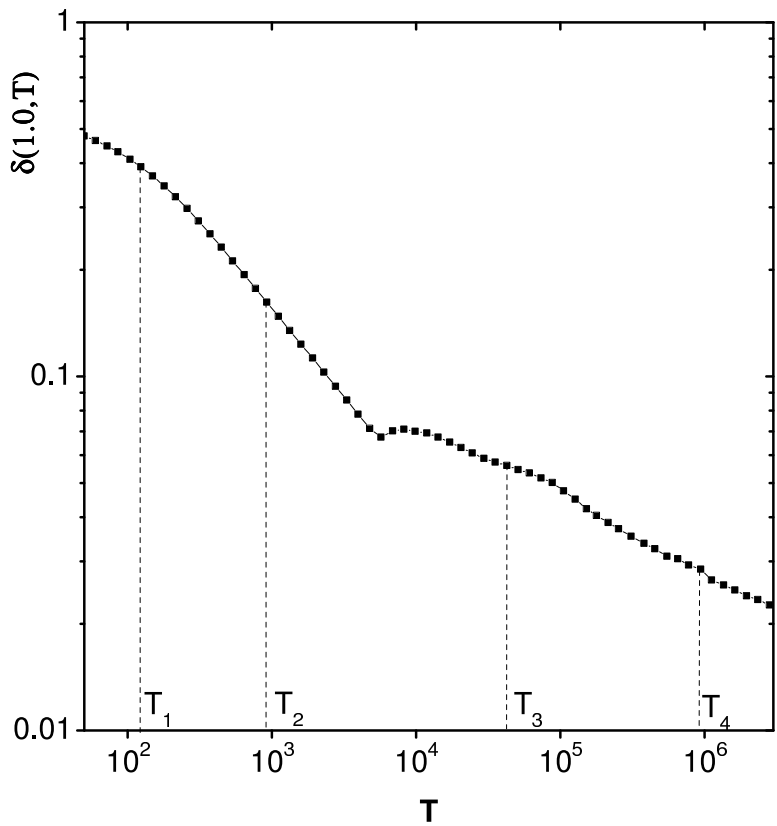}
\caption{The behavior of $\delta (1.0, T)$ for the spatio-temporal dynamics shown
in Figure~\ref{snapshot0}. The presence of two distinct stages of relaxation can 
clearly be observed. }
\label{delta}
\end{figure}

\begin{figure}
\includegraphics[width=4in]{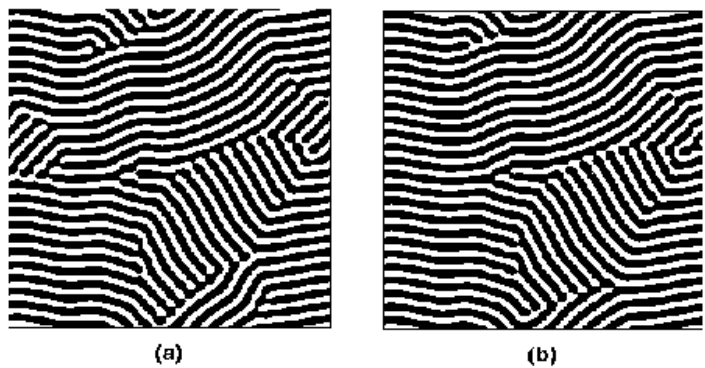}
\caption{Sharp reductions in $\delta (\beta, T)$ (similar to that near $T_4$ in 
Figure~\ref{snapshot0}) correspond to the disappearance of domains or defects. The images 
(a) before, and (b) after such an event. It is found that $\sigma (\beta, T)$
fails to be monotonically decreasing during and immediately following such events.}
\label{defects}
\end{figure}

\begin{figure}
\includegraphics[width=5.5in]{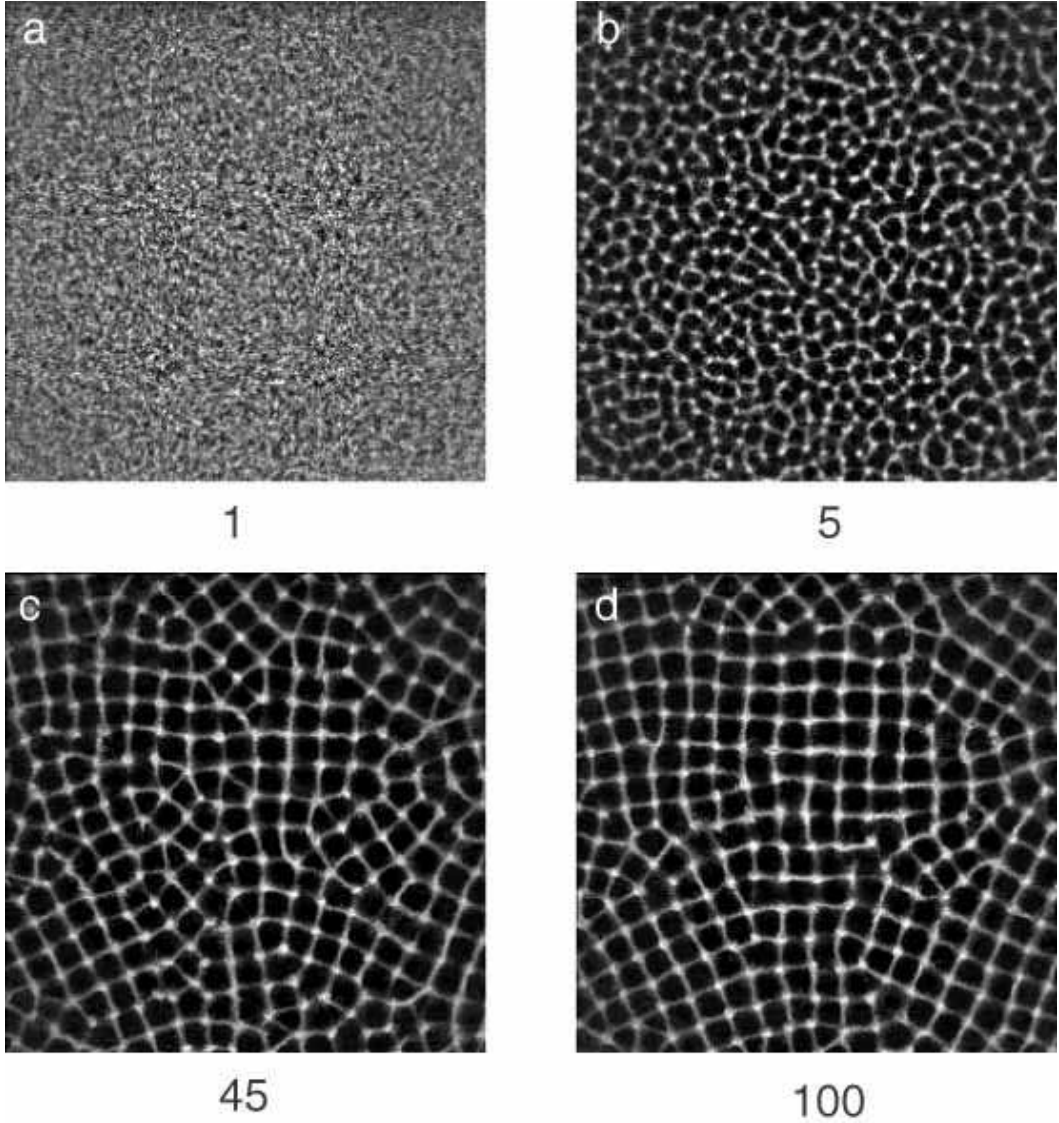}
\caption{the time evolution of an oscillated layer of grains following a
rapid change in peak acceleration of the container from $2.2g$ to $3.0g$.
Below the onset of patterns (which occurs at approximately $2.5g$), the 
layer is featureless but noisy while above onset defect-free square patterns
are stable. The numeral below each snapshot indicates the number of 
layer-plate collisions after the last layer takeoff before the acceleration was changed. The oscillation 
frequency in this experiment is $f=27$ Hz.}
\label{snapshot_exp}
\end{figure}

\begin{figure}
\includegraphics[width=4in]{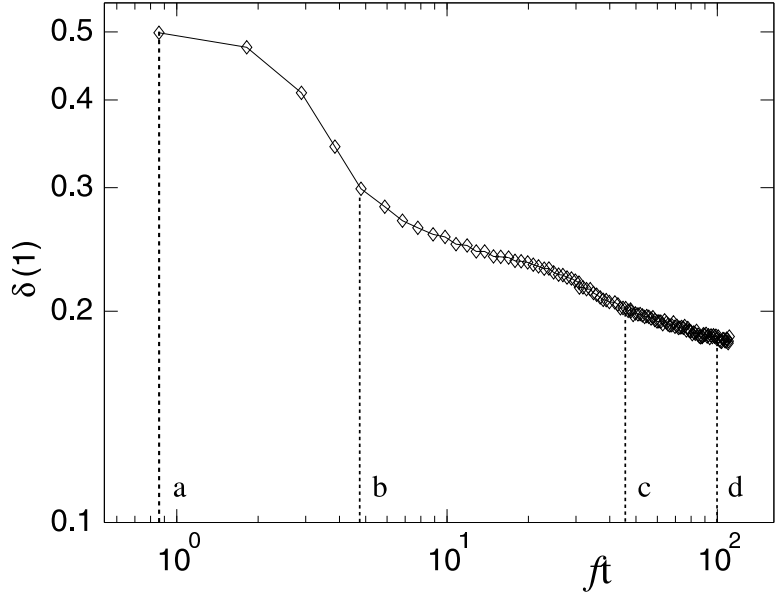}
\caption{The time evolution of $\delta(1.0)$ for the sequence shown in 
Figure~\ref{snapshot_exp}. The abscissa is in units of the number of container oscillations
from the last layer takeoff take before $\Gamma$ was changed. The dashed lines indicate the time that snapshots 
in Figure~\ref{snapshot_exp} were taken. Two distinct stages of pattern evolution are 
clearly visible; the initial rapid decay is consistent with $t^{-0.5}$,
while the slower decay scales on average like $t^{-0.12}$. However, it is 
difficult to differentiate stages II and III from the dynamics of a single
moment of the disorder function.}
\label{delta_exp}
\end{figure}

\begin{figure}
\includegraphics[width=4in]{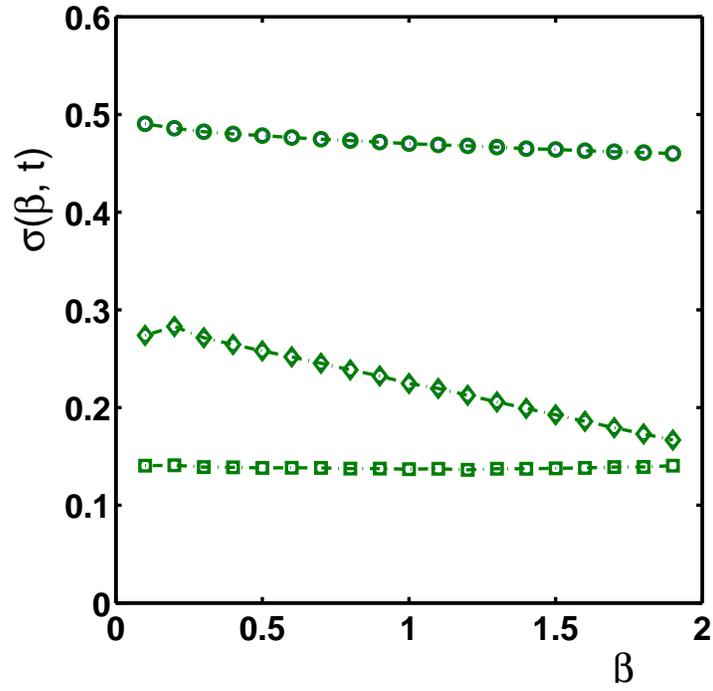}
\caption{$\sigma(\beta, T)$ calculated for patterns shown in Figure~\ref{snapshot_exp}.
Results from stages I, II, and III are shown in circles, 
diamonds and squares respectively. $\sigma (\beta, t)$ for the three stages were calculated
using intervals shown in Figure~\ref{snapshot_exp} (3-5 oscillations, 40-50 oscillations, 
and 85-105 oscillations following the initiation of the experiment). Now 
stages II and III can clearly be delineated.}
\label{expt}
\end{figure}

\begin{figure}
\includegraphics[width=5.5in]{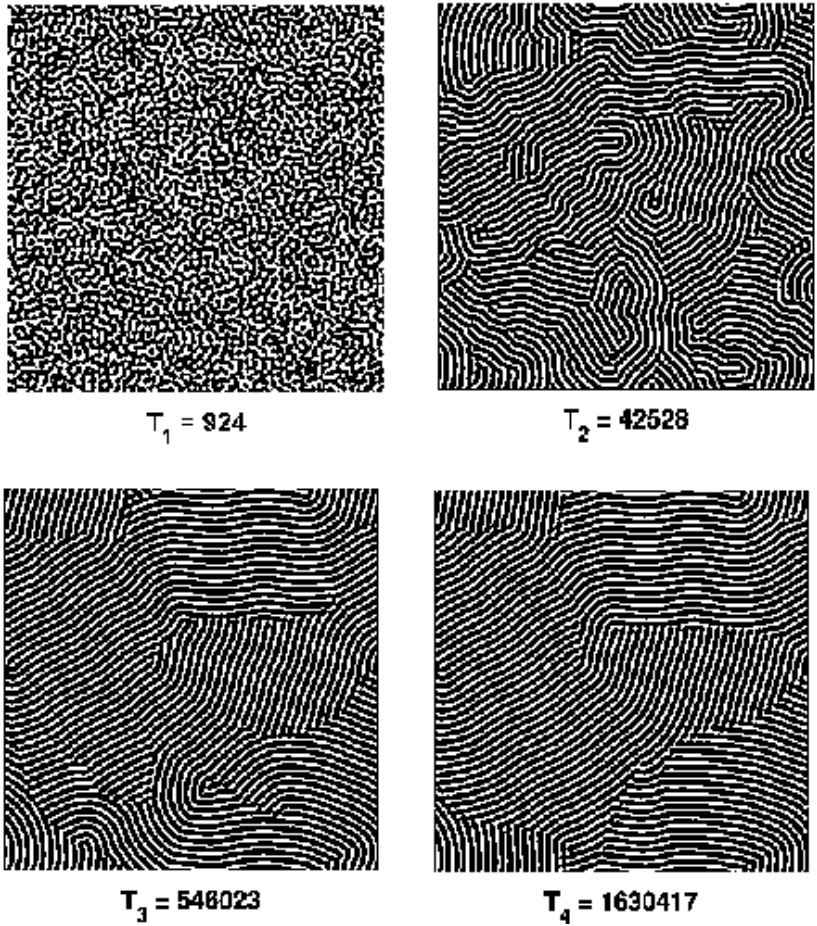}
\caption{Four snapshots of the stochastic spatio-temporal dynamics of a random initial state
($|u({\bf x}, t)| < 0.001$) under Eqn.~\ref{she} with $\epsilon=0.01$, $k_0=3$,
and $F=0.0001$. The $T_1=924$ ls in stage I, $T_2=45,528$ in stage II,  
and the last two $T_3=546,023$ and $T_4=1,630,417$ are in stage III. 
Note that the pattern continues to evolve, albeit slowly, in stage III.}
\label{snapshot}
\end{figure}

We set out to compare and contrast pattern dynamics in distinct systems.
It is important to note that
such a task can only be achieved using ``configuration independent" 
characteristics; i.e., measures that depend on control parameters of the
underlying system, and not on the (random) initial states.
For complex patterns with local planforms such
as stripes, squares or hexagons (e.g., Figures~\ref{snapshot0}, \ref{snapshot_exp} and \ref{snapshot}),
such measures include the width of the structure factor~\cite{eldAvin,croAmei},
defect density~\cite{houAsas}, length of domain boundaries~\cite{ouyAswi2},
and the disorder function~\cite{gunAjon,gunAhof}. 

Analysis of patterns generated in the Swift-Hohenberg equation and on a vibrated 
layer of brass beads show that during early times all components of
the disorder function relax as $t^{-0.5}$. We showed that the index can be obtained
by noting that during this stage the intensity of the field representing the pattern
is sufficiently small and hence the spatio-temporal dynamics can be considered 
to be linear. The saturation of the field is followed, in each case, by a second stage
where the relaxation is non-universal and the growth indices $\sigma (\beta, t)$ are 
$\beta$ dependent; the latter implies that the growth of distinct features of  
patterns occur at different rates.  
Experimental patterns exhibit a third stage at times very long compared
to the vibrational period of the container~\cite{golAsha}, where $\sigma (\beta, t)$
is once again independent of $\beta$. It is not possible to differentiate
it from stage II by analyzing the behavior of a single index; rather, it requires the use of 
the entire family of indices $\sigma(\beta, t)$. Unlike in stage I, 
the value of $\sigma (\beta, t)$ is not universal. Stage III is not present in 
spatio-temporal dynamics under the non-stochastic Swift-Hohenberg equation, even when
non-variational terms and finite-size effects are included. Our work suggests
that the ingredient required to initiate stage III is stochasticity of
the underlying spatio-temporal dynamics. 

It should be noted that $\delta (\beta, t)$ was constructed to quantify 
disorder in patterns that consist of local planforms which satisfy the Helmholtz 
equation~\cite{gunAhof}. Structures formed in other systems, such as in the
X-Y model~\cite{bray}, in spinodal decomposition~\cite{cahAhil},
or in epitaxial growth~\cite{vveAzan} do not have such local planforms.
Measures to conduct irreversible statistical analyses of such structures
remain to be identified.

The authors would like to acknowledge discussions with 
M. Shattuck. This research is supported by grants from the 
National Science Foundation (DJK, GHG), the R. A.. Welch 
Foundation (DJK), and the Office of Naval Research (GHG).

\begin{thebibliography}{99}
\bibitem{footnote0} Methods used to integrate these spatio-temporal dynamics
are described in Ref.~\cite{gunArat}.
\bibitem{footnote1} These integrals cannot be calculated due to the nonanalyticity 
of $|x|$. The values for the analytic case, $\beta=2$, are easily obtained 
using Parseval's theorem.
\end{thebibliography}
\end{document}